\documentclass[pra,aps,twocolumn]{revtex4}
\usepackage{amssymb}
\usepackage{latexsym}
\usepackage{endnotes}
\usepackage{graphicx}
\usepackage{epsf}
\newcommand{\Z}{{\mathbb Z}}

\newcommand{\R}{{\mathbb R}}

\def\be{\begin{equation}}
\def\ee{\end{equation}}
\def\bea{\begin{eqnarray}}
\def\eea{\end{eqnarray}}

\def\d{{\,\rm d}}

\def\vfi{\varphi}
\def\fihat{\hat{\varphi}}
\def\0{{\bf 0}}
\def\a{{\bf a}}
\def\b{{\bf b}}
\def\k{{\bf k}}

\def\q{{\bf q}}
\def\n{{\bf n}}
\def\r{{\bf r}}
\def\p{{\bf p}}

\def\x{{\bf x}}
\def\y{{\bf y}}
\def\K{{\bf K}}

\def\Rr{{\bf R}}

\def\h2m{\frac{\hbar^2}{2m}}
\def\p0{{P_{\beta H^0_N}}}

\begin{document}

\title{{\flushleft{\small {\rm Published: 
Phys. Rev. B {\bf 74}, 104117 (2006)}\\}}
\vspace{1cm}
\large\bf From bcc to fcc: interplay between oscillating long-range and repulsive short-range
forces
}
\author{Andr\'as S\"ut\H o\\
Research Institute for Solid State Physics and Optics,
Hungarian Academy of Sciences\\
P. O. B. 49, H-1525 Budapest, Hungary\\
E-mail: suto@szfki.hu}
\thispagestyle{empty}
\begin{abstract}
\noindent
This paper supplements and partly extends an earlier publication, Phys. Rev. Lett. {\bf 95}, 265501
(2005). In $d$\,-dimensional continuous space
we describe the infinite volume ground state configurations (GSCs) of
pair interactions $\vfi$ and $\vfi+\psi$,
where $\vfi$ is the inverse Fourier transform of a nonnegative function vanishing
outside the sphere of radius $K_0$, and $\psi$ is any nonnegative finite-range interaction of range
$r_0\leq\gamma_d/K_0$, where $\gamma_3=\sqrt{6}\pi$.
In three dimensions the decay of $\vfi$ can be as slow as $\sim r^{-2}$,
and an interaction of asymptotic form $\sim\cos(K_0r+\pi/2)/r^3$ is among the examples.
At a dimension-dependent density $\rho_d$ the ground state of $\vfi$ is a unique Bravais lattice,
and for higher densities it is continuously degenerate: any union
of Bravais lattices whose reciprocal lattice vectors are not shorter than $K_0$ is a GSC.
Adding $\psi$ decreases the ground state degeneracy which, nonetheless, remains continuous in the
open interval $(\rho_d,\rho_d')$, where $\rho_d'$ is the close-packing density of hard balls
of diameter $r_0$. The ground state
is unique at both ends of the interval. In three dimensions this unique
GSC is the bcc lattice at $\rho_3$ and the fcc lattice at $\rho_3'=\sqrt{2}/r_0^3$.

\vspace{2mm}
\end{abstract}
\maketitle

\section{Introduction}

In an earlier Letter \cite{Su} we described the infinite volume
ground state configurations (GSCs) of a
class of classical particle interactions in $d\geq 1$ dimensional continuous space. These pair
interactions have a nonnegative
Fourier transform vanishing above some finite wave number, $K_0$. We
proved that at a threshold density $\rho_d\propto K_0^d$ there is
a unique periodic GSC (the basic-centered cubic lattice in three dimensions),
and above $\rho_d$ the set of
GSCs is continuously degenerate and contains periodic and aperiodic configurations. While this
was probably the first result providing specific examples in three (and higher) dimensions,
important rigorous work preceded it in lower dimensions; see, for instance, Kunz~\cite{Kunz}
on the one-dimensional one-component plasma, Ventevogel, Nij\-boer
and Ruijgrok~\cite{VNR} and Radin~\cite{Rad1} on ground states in one dimension
and Theil's recent proof of ground state crystallization in two dimensions \cite{The}.
Although we will be concerned only with ground state ordering, let us note that
rigorous results on phase transitions or ordering in a continuum at positive
temperatures do not abound, and all are about more or less contrived model systems.
The one-component plasma in one dimension is ordered at all
temperatures~\cite{Kunz}; Ruelle~\cite{Rue} proved segregation in a two-component system (the
Widom-Rowlinson model);
Lebowitz, Mazel and Presutti~\cite{LMP} proved vapor-liquid transition for particles with
two-body attractive and
four-body repulsive interactions near the mean-field limit;
in a one-dimensional model with an unstable interaction low-temperature
freezing into an ordered configuration was shown by the present author~\cite{Su2}; and recently
Bowen \emph{et al.} proved fluid-solid phase transition in a two-dimensional system of decorated
hard hexagons~\cite{BLRW}.

In this paper we supplement and partly extend the
results of \cite{Su}. Apart from recalling the definitions, we do not repeat what
is written there. The main part of the theorem of \cite{Su} will be stated and proven
in a new, simpler form, emphasizing the nice algebraic structure of the set of GSCs.
Moreover, the results
will be extended to interactions of a non-integrable decay.
The proof in \cite{Su} was based on the Poisson summation formula. This formula is
widely used in physics; one of its earliest and most famous applications was
the calculation of the Madelung constant of ionic crystals by Ewald \cite{Ew}.
The formula involves at least one infinite summation,
and neither the convergence of the infinite sum(s) nor the equality
of the two sides is guaranteed. Although the results of \cite{Su} were already
formally valid to the
larger class of interactions, we stated them only for a restricted class, those of the
strongly tempered interactions (see later), because no argument supporting the applicability
of the Poisson formula to functions of a non-integrable decay was given in that paper.
The extension of this formula constitutes
an active field of research in mathematics,
see e.g.~\cite{DEK}, but is not our main concern here. Therefore,
without looking for the most general formulation, we
propose an extension just suitable for our purposes.
As a matter of fact, the extension involves also
the notion of a ground state configuration. The definition of a GSC is based on infinite
sums that are absolutely convergent for strongly tempered interactions, but
only conditionally convergent for interactions of a non-integrable decay, and the way they
converge has to be specified.

Another, gratuitous,
extension, mentioned but not exploited in \cite{Su}, will be obtained by modifying
the short-range behavior of the interaction. The inverse Fourier transform $\vfi$ of an integrable
function is bounded and continuous | this is our case. Such bounded functions
play a role as soft effective interactions in polymer physics \cite{KL}, but not in traditional
solid state physics
where Pauli exclusion gives rise to a practically infinite repulsion at overlaps of atoms.
Imagine, however,
that an infinite configuration $X$ was shown to be a GSC of $\varphi$. Then $X$
will be a GSC of all interactions $\varphi+\psi$, where $\psi$ is non-negative and vanishes at
and above the nearest-neighbor distance of $X$: $\psi$ does not contribute to the specific energy
of $X$, and can only increase the energy of any perturbation of $X$. Reversing the argument,
we may start with $\varphi+\psi$, where $\psi$ is of bounded support, non-negative and may
contain a hard core or diverge at the origin as fast as we wish. Then $\varphi+\psi$ has
common GSCs with $\varphi$ if the support of $\psi$ is small enough and the density is not too
high.

The physical importance of the above two extensions is that with them we obtain the GSCs
of interactions whose asymptotic form is $\sim\cos K_0r/r^3$, as that of the RKKY interaction,
but which can be arbitrarily strongly repulsive at small distances.
As noted also by Likos \cite{Lik}, such interactions can model
those between ions in metals and be relevant in the explanation of
the crystal structure of certain metals. However, further study is necessary before
any conclusion could be drawn about this question. As an immediate gain, we will find that
at some density
$\rho_3'>\rho_3$ the unique GSC of $\vfi+\psi$ is the fcc lattice, while at $\rho_3$
it is the bcc lattice. This transition from bcc to fcc with an increasing density
is the consequence of an interplay between a long-range oscillating interaction
(which is short-range in Fourier space) and a short-range positive pair potential.

The following section is the central part of the paper. After introducing the necessary
definitions we enounce a theorem in a rather compact form, and then expand its content in a
series of remarks. The Poisson summation formula is presented here as a lemma. In Section III
we prove the lemma, an auxiliary statement about Bravais lattices, and the theorem. This
section also contains the proof of a
general assertion about the non-existence of metastable ground states.
The paper is closed with a brief Summary.

\section{Definitions, notations, results}

We consider a system of identical classical particles in $\R^d$, that interact through translation
invariant symmetric pair interactions, $\vfi(\r-\r')=\vfi(\r'-\r)$.
Rotation invariance is not supposed.
An $N$-particle configuration ($N\leq\infty$)
is a sequence $(\r_1,\ldots,\r_N)$ of $N$ points of $\R^d$ and
will be denoted by $B$ (referring always to a Bravais lattice),
$R$, $X$ and $Y$. While the order of the points
is unimportant,
two or more particles may coincide in a point, resulting $\r_{i_1}=\cdots=\r_{i_m}$.
Such a coincidence can occur if $\vfi(\0)$ is finite, and it indeed occurs in certain GSCs to
be described below. Throughout the paper, the notation $\varphi$ will be reserved to bounded
interactions; unbounded interactions, such as those
diverging at the origin or including a hard core, will
be composed as $\varphi+\psi$.
The number of points in
$R$ will be denoted by $N_R$. The energy of a finite configuration $R$ is
\be\label{U1}
U(R)=\frac{1}{2}\sum_{\r,\r'\in R, \r\neq\r'}\vfi(\r-\r').
\ee
Let $R$ be a finite and $X$ be an arbitrary configuration. The interaction energy of $R$ and $X$
is
\be\label{int}
I(R,X)=\sum_{\r\in R}I(\r,X)=\sum_{\r\in R}\sum_{\x\in X}\vfi(\r-\x),
\ee
and the energy of $R$ in the field of $X$ is
\be\label{rel}
U(R|X)=U(R)+I(R,X).
\ee
If $X$ is an infinite configuration, the infinite sum in (\ref{int}) has to be convergent.
This imposes conditions on both $X$ and $\vfi$, and the stronger the condition on $X$, the
weaker it can be on $\vfi$. For example, one may ask $I(R,X)$ to be finite for
\emph{every} $X$ that is locally uniformly finite, meaning the existence
of an integer $m_X$ such that
the number of particles in a unit cube everywhere in $\R^d$ stays below $m_X$.
This was our choice in \cite{Su}; the corresponding condition on the interaction is
strong temperedness which for a bounded $\vfi$ reads
\be\label{abssum}
\sum_{\x\in X}|\varphi(\x)|<\infty
\ee
for any locally uniformly finite $X$.

\noindent
\emph{Definition.}|
Given a real $\mu$, $X$ is a ground state configuration of $\vfi$
for chemical potential $\mu$ (a $\mu$GSC) if for any bounded domain $\Lambda$
and any configuration $R$
\be\label{GSC'}
U(R\cap\Lambda|X\setminus\Lambda)-\mu N_{R\cap\Lambda}
\geq U(X\cap\Lambda|X\setminus\Lambda)-\mu N_{X\cap\Lambda}
\ee
where $X\cap\Lambda$ and $X\setminus\Lambda$ are parts of $X$
inside and outside $\Lambda$, respectively. $X$ is a
ground state configuration (GSC)
if (\ref{GSC'}) holds true for every $R$ such that $N_{R\cap\Lambda}=N_{X\cap\Lambda}$.

A seemingly more general, but actually equivalent definition is as follows.
$X$ is a $\mu$GSC (respectively, $X$ is a GSC) if for any finite part $X_f$ of $X$ and any
finite $R$ (respectively, any $R$ such that $N_R=N_{X_f}$)
\be\label{GSC}
U(R|X\setminus X_f)-\mu N_{R}\geq U(X_f|X\setminus X_f)-\mu N_{X_f}\ .
\ee

If $\vfi$ is strongly tempered, we can | at least in principle |
test any locally uniformly finite $X$ to be, or not,
a GSC according to (\ref{GSC}).
Ground states of interactions that violate condition (\ref{abssum}),
as those between ions in metals mediated by the
Friedel oscillation of the conduction electrons,
can be defined only within a more restricted set of configurations.
Intuitively, ground states cannot be arbitrary sets of points, they are arrangements
with some good averaging
(ergodic) property. Specifically, we shall look for them only among
periodic configurations and their unions. Simultaneously, the infinite sums appearing in
(\ref{rel}) will be suitably interpreted.

A Bravais (direct) lattice $B=\{\sum_{\alpha=1}^dn_\alpha \a_\alpha | \n\in\Z^d\}$
is regarded as an infinite
configuration. Here $\a_\alpha$ are linearly independent vectors and
$\n=(n_1,\ldots,n_d)$ is a $d$-dimensional integer. The dual (reciprocal) of $B$ is
the Bravais lattice
$B^*=\{\sum n_\alpha\b_\alpha | \n\in\Z^d\}$ where
$\a_\alpha\cdot\b_\beta=2\pi\delta_{\alpha\beta}$.
The nearest neighbor distances
in $B$ and $B^*$ are denoted by $r_B$ and
$q_{B^*}$, respectively.
The latter is related to the density of $B$ via $\rho(B)=c_{``B"} (q_{B^*})^d$,
where $c_{``B"}$ is determined by the aspect ratios and angles of the primitive cell of $B$.
[Notational remark: $B$ and $B^*$ will always refer to specific Bravais lattices as given above.
$``B"$ refers to the family of all Bravais lattices of the type of $B$,
characterized by dimensionless quantities. $``B"$
may take on the `value' bcc, fcc, simple cubic, and so on.]
We shall look for GSCs of the form
$
X=\cup_{j=1}^J(B_j+\y_j)
$
where $B_j$ are Bravais lattices and $B_j+\y_j$ is $B_j$ shifted by the vector $\y_j$.
We shall refer to configurations of this form as unions of periodic configurations.
If $J=1$, $X$ is a Bravais lattice. If $B_j=B$ for each $j$ then $X$ is periodic. If at least
two different Bravais lattices are involved in the union then $X$ is
either periodic or aperiodic, and may contain overlapping points that are to
be counted with repetition. The density of $X$ is $\rho(X)=\sum_{j=1}^J\rho(B_j)$. Now
\be\label{IxX}
I(\r,X)=\sum_{j=1}^J I(\r-\y_j,B_j),
\ee
so the sum to be interpreted is
\be\label{inter}
I(\r,B)=\sum_{\Rr\in B}\vfi(\r+\Rr)
=\sum_{\n\in\Z^d}\vfi\left(\r+\sum_{\alpha=1}^dn_\alpha\a_\alpha\right).
\ee
The interaction $\vfi$ will be defined as the inverse Fourier transform of a function
$\fihat\in L^1(\R^d)$ that vanishes outside the ball of radius $K_0$:
$
\vfi(\r)=(2\pi)^{-d}\int_{k<K_0}\fihat(\k)e^{i\k\cdot\r}\d\k.
$
Then $\vfi$ is continuous and all its derivatives exist and are also
continuous functions decaying at infinity; in fact,
$\vfi(\r)$ is an entire function of $\r$ \cite{rem1}.
By definition,
\be\label{Poisson-Abel}
\sum_{\Rr\in B}\vfi(\r+\Rr)=\lim_{\varepsilon\downarrow 0}
\sum_{\Rr\in B}e^{-\varepsilon|\r+\Rr|^2}\vfi(\r+\Rr)
\ee
provided that the limit exists. In our case a weaker, e.g. exponential, tempering would
suffice and give the same result;
the Gaussian tempering is more convenient to work with in an arbitrary dimension.
Note that the sum on the right-hand side is absolutely convergent
for any $\varepsilon>0$. It is easily seen that whenever the sum in (\ref{inter})
is absolutely convergent, Eq.~(\ref{Poisson-Abel}) yields the same result.
This is the case of all the examples given in~\cite{Su}. In the absence of absolute convergence,
the sum (\ref{inter}) can still be conditionally convergent; e.g.
with some mild additional assumption on $\fihat$ one can show that
\be\label{limsum}
I(\r,B)=\lim_{N_1,\ldots,N_d\to\infty}\sum_{\n\in\Z^d, |n_\alpha|<N_\alpha}
\vfi\left(\r+\sum_{\alpha=1}^dn_\alpha\a_\alpha\right)
\ee
exists and agrees with the result suggested by the Poisson summation formula.
However, the proof of this formula is simpler with the definition (\ref{Poisson-Abel}).

THEOREM. Let $\fihat\in L^1(\R^d)$ be a real function with the following properties:

\noindent
(1) $\fihat$ is continuous at the origin,\\
(2) $\fihat(-\k)=\fihat(\k)$,\\
(3) $\fihat\geq 0$ and\\
(4) there is some $K_0$ such that $\fihat(\k)=0$ for $|\k|>K_0$.

(i) Define
$
\vfi(\r)=(2\pi)^{-d}\int\fihat(\k)e^{i\k\cdot\r}\d\k
$.
Choose Bravais lattices $B_1,\ldots,B_J$ such that each $q_{B_j^*}\geq K_0$,
where equality is allowed only if $\fihat$ is continuous at $|\k|=K_0$.
Then $X=\cup_{j=1}^J(B_j+\y_j)$ is a GSC of $\vfi$ for arbitrary translations $\y_j$
and it is also a $\mu$GSC for
$\mu=\rho(X)\fihat(\0)-\frac{1}{2}\vfi(\0)$. The energy per unit volume of $X$ is
$e(X)=\epsilon\textbf{(}\rho(X)\textbf{)}$ where
\be
\epsilon(\rho)=\frac{1}{2}\rho[\rho\fihat(\0)-\vfi(\0)]
\ee
is the minimum of the energy density among unions of periodic
configurations of density $\rho$. GSCs of the above properties exist in a semi-infinite
density interval $[\rho_d,\infty)$.

(ii)
If $\vfi$ is strongly tempered and $X$ is locally uniformly finite with existing
$\rho(X)\geq\rho_d$ and $e(X)>\epsilon\textbf{(}\rho(X)\textbf{)}$, then $X$ is not a GSC. If
$\vfi$ is not strongly tempered but the limit (\ref{limsum}) on Bravais lattices exists, then
any union $X$ of
periodic configurations with $\rho(X)\geq\rho_d$ and $e(X)>\epsilon\textbf{(}\rho(X)\textbf{)}$
is not a GSC.

(iii) Let $r_0\leq \gamma_d/K_0$, where $\gamma_1=2\pi$, $\gamma_2=4\pi/\sqrt{3}$ and
$\gamma_3=\sqrt{6}\pi$, and let
$\psi$ be a real function such that
$\psi(\r)\in [0,\infty]$ and $\psi(\r)=0$ for $r\geq r_0$. If $r_{B}\geq r_0$ and
$q_{B^*}\geq K_0$, then $B$ is a GSC and a $\mu$GSC of $\vfi+\psi$ with $\mu$ and $e(B)$
given above, not depending on $\psi$. GSCs of the above properties exist in a
density interval $[\rho_d,\rho_d']$.

\emph{Remarks.}|
1. Compared with the theorem of \cite{Su}, the conditions on $\vfi$ are formulated
uniquely via $\fihat$, and are considerably weaker. For instance, $\fihat$
or its derivative can be discontinuous at $K_0$. Here are two examples in three dimensions:
$\fihat(\k)\equiv 1$ for $k<K_0$ yields ($k=|\k|$, $r=|\r|$)
\be\label{ex1}
\vfi(\r)=-(K_0/2\pi^2)\cos K_0r/r^2+(1/2\pi^2)\sin K_0r/r^3\ ,
\ee
while with $\fihat(\k)=1-k/K_0$ for $k<K_0$ we obtain
\be
\vfi(\r)=\frac{\cos(K_0r+\pi/2)}{2\pi^2r^3}+\frac{1-\cos K_0r}{\pi^2K_0r^4}\ .
\ee
Both are conditionally summable on Bravais lattices [if $B^*$ has no point on the
sphere $|\K|=K_0$,
in the case of (\ref{ex1})], as defined in Eq.~(\ref{limsum}).

2. We proved in \cite{Su} that the condition $q_{B^*}\geq K_0$ can
be satisfied only if $\rho\geq\rho_d$,
a dimension-dependent threshold density at which $q_{B^*}=K_0$ for
a unique Bravais lattice $B$, and this is the unique periodic GSC;
in particular, $\rho_3=K_0^3/8\sqrt{2}\pi^3$
and the lattice is the bcc one. The above form of the theorem shows that for
$\rho_d\leq \rho<2\rho_d$ no union is available, only Bravais lattices can be GSCs.
In \cite{Su} we gave also the densities of some Bravais lattices $B$
at which $q_{B^*}=K_0$. Recalling these values,
\bea\label{rhos}
\rho_{\rm bcc}=\rho_3
<\rho_{\rm fcc}=\frac{4\sqrt{2}}{3\sqrt{3}}\,\rho_3=1.089 \rho_3\nonumber\\
<\rho_{\rm sh}=\sqrt{\frac{3}{2}}\,\rho_3  %=1.225\rho_3
<\rho_{\rm sc}=\sqrt{2}\rho_3 %=1.414\rho_3
\eea
(sh = simple hexagonal with $c/a=\sqrt{3}/2$,
sc = simple cubic), one can see that all the high-symmetry
Bravais lattices appear as GSCs between $\rho_3$ and $2\rho_3$.
Also, if $\rho(Z)$ denotes the density of a metal of valency $Z$
then, in the free-electron approximation and supposing a spherical Fermi surface of radius
$k_F=K_0/2$,
$\rho(Z)=(\sqrt{2}\pi/3Z)\rho_3=(1.481/Z)\rho_3$ which for
$Z=1$ is in this interval.
In general, in the interval
$n\rho_d\leq\rho<(n+1)\rho_d$ the ground state configurations are unions of at most $n$
Bravais lattices, each of density $\geq\rho_d$. Thus, the simplest aperiodic GSCs,
unions of two incommensurate Bravais lattices, appear only if $\rho\geq 2\rho_d$.
For example, in 3 dimensions at $2\rho_3$ they are the unions of two bcc lattices rotated
and possibly shifted with respect to each other.

3. The family of all the GSCs of $\vfi$ above the density $\rho_d$
is closed on unions. This is obvious from the present
formulation, because
the union of two GSCs of the form given in the theorem is a configuration of the same form,
so it is necessarily also a GSC.
Recall from \cite{Su} that a periodic configuration $X$ is called
$B$-periodic if $X=\cup_{j=1}^J(B+\y_j)$ and $B$ is chosen so as to minimize $J$.
If $X_m$ are $B_m$-periodic configurations then $\cup X_m$ is periodic if and only if
$B=\cap B_m$ is a $d$-dimensional Bravais lattice. Because $B\subseteq B_m$,
$B^*\supseteq B_m^*$ and $q_{B^*}\leq q_{B_m^*}$. It follows that
a $B$-periodic configuration $X=\cup_{j=1}^J(B+\y_j)$
can be a GSC even if $q_{B^*}<K_0$, provided that it can be written
also as $X=\cup_{j=1}^{J'}(B_j+\y'_j)$, where $q_{B_j^*}\geq K_0$ for $j=1,\ldots,J'$.
This means that on average the $B_j$s are denser than $B$ and thus $J'<J$. Logically, if we
permit different Bravais lattices to occur in the union forming a periodic configuration,
the number of components may be decreased.

4. A $\mu$GSC is, by definition, also a GSC because it satisfies a stronger condition.
Therefore, in the theorem
it would have been enough to say that $X$ or $B$ is a $\mu$GSC. We wanted to emphasize that
the theorem strengthens
that of \cite{Su} by stating that the opposite is also true: a GSC
is always a $\mu$GSC, even if $\fihat(\0)=0$ and thus $\vfi$ is not superstable. In the latter
case $\mu=-\frac{1}{2}\vfi(\0)$, independently of the density of the ground state configuration.

5. Assertion (ii) of the Theorem
extends to a larger class of configurations and pair potentials earlier results by
Sewell \cite{Sew} and Sinai \cite{Sinai} on the absence of metastability
for strongly tempered interactions. Following the usual definition \cite{Sinai},
we apply the term ``ground state configuration'' as a synonym of a locally stable configuration.
Thus, a GSC could be globally unstable, meaning that by some perturbation involving
infinitely many particles its energy density could be decreased. Such a GSC might be called
metastable. However, we have to precise the kind of infinite perturbations we allow. If the
particle density is allowed to vary, usually the absolute minimum of the energy density is
attained at a single value of $\rho$, and all GSCs of a different density should be considered
metastable. In this sense, the unique stable GSC of an everywhere
positive interaction is the vacuum, and
for the interactions $\vfi$ studied in this paper the globally stable GSCs are
the $\mu$GSCs belonging to $\mu=0$
(hence, to $\rho=\vfi(\0)/2\fihat(\0)$, if this value is finite and not smaller than $\rho_d$).
We adopt a more restrictive definition of metastability, not allowing the density to vary. Then
the configurations characterized by the theorem as GSCs are
not metastable because their energy density is the attainable minimum for their density, and
no other configuration (within the specified class) can be a metastable ground state.

6. A sufficient condition for
$\vfi$ to be strongly tempered is that $|\vfi(\r)|\leq Cr^{-d-\eta}$ for $r>r'$, where
$C$, $\eta$ and $r'$ are some positive numbers. In our case this holds, for example, if
besides conditions (1)-(4), $\fihat$ is 3 times differentiable, see Eq.~(9) of
\cite{Su}.

7. Point (iii) of the theorem needs more an explanation than a formal proof.
When we ask $r_B\geq r_0$, we limit the role of $\psi$
to reducing the degeneracy of the GSCs of $\vfi$. The largest allowed range of $\psi$,
$r_0=\gamma_d/K_0$, equals the nearest-neighbor distance of the unique
GSC of $\vfi$ at the density $\rho_d$: the uniform chain, the triangular lattice and the bcc
lattice for $d=1$, 2 and 3, respectively.
The theorem makes no prediction if $r_0> \gamma_d/K_0$, because no Bravais
lattice satisfies both conditions $r_{B}>\gamma_d/K_0$ and $q_{B^*}\geq K_0$.
This is an obvious consequence of $r_Bq_{B^*}=2\pi$ in one dimension, and of
\be\label{max}
\gamma_d=\max_B r_Bq_{B^*}=\left\{\begin{array}{ll}
                         4\pi/\sqrt{3} &(d=2)\nonumber\\
                         \sqrt{6}\pi   &(d=3),
                         \end{array}\right.
\ee
with the maximum
attained if $B$ is the triangular lattice in two, and the bcc (or fcc) lattice in three
dimensions. We prove (\ref{max}) in the next section, and suppose henceforth that
$r_0K_0\leq\gamma_d$.
For a given lattice type $``B"$, $r_Bq_{B^*}=\gamma_{``B"}=\gamma_{``B^*"}$,
independent of the density.
This implies that the simultaneous inequalities
$r_B\geq r_0$ and $q_{B^*}\geq K_0$ hold in a (closed) interval of the
density, whose lower and upper boundaries are implicitly determined
by $q_{B^*}=K_0$ and $r_B=r_0$,
respectively. If $\rho$ is in this ``stability interval" $I_{``B"}$,
then $B$ is a GSC of $\vfi+\psi$, with energy density
$e(B)=\epsilon(\rho)$.
If $\rho$ is not in $I_{``B"}$, then $e(B)>\epsilon(\rho)$
[provided the strict positivity of $\fihat$ for $k<K_0$ and of
$\psi$ for $r<r_0$, that we suppose now], and $B$ is surely not a GSC of $\vfi+\psi$ if
$\rho$ is still in the stability interval of some other Bravais lattice $\tilde{B}$: In this case
$\tilde{B}$ is a GSC with energy density $e(\tilde{B})=\epsilon(\rho)$, and due to the absence
of metastability, $B$ cannot be a GSC.
$I_{``B"}$ shrinks to a single point if $r_0=\gamma_{``B"}/K_0$
and disappears if $r_0>\gamma_{``B"}/K_0$.
In two dimensions $\gamma_2=\gamma_{\rm tr}$, so this
may happen with any Bravais lattice other than the triangular one. Because
the triangular lattice is self-dual, at a given density it has the largest $q_{B^*}$ \emph{and}
the largest $r_B$ among the two-dimensional Bravais lattices. Therefore its stability interval
\be
I_{\rm tr}=[\rho_2,\rho(r_{\rm tr}=r_0)]\equiv[\rho_2,\rho_2']=
\left[\frac{\sqrt{3}K_0^2}{8\pi^2},\frac{2}{\sqrt{3}r_0^2}\right]
\ee
contains in its interior the stability intervals of all the other Bravais lattices.
Figure 1 shows the stability intervals of the triangular and the square lattices for all allowed
values of $r_0$. If $\rho$
falls into $I_{\rm tr}$ but outside $I_{``B"}$, then $B$ is certainly not a GSC
of $\vfi+\psi$ at this density.
In particular, at the two ends of $I_{\rm tr}$ the only GSC is the triangular lattice. The
uniqueness at
the upper value $\rho_2'=2/\sqrt{3}r_0^2$ is new, and is due to $\psi$.

\begin{figure}[htb]
\centerline{\includegraphics[width=8.5cm]{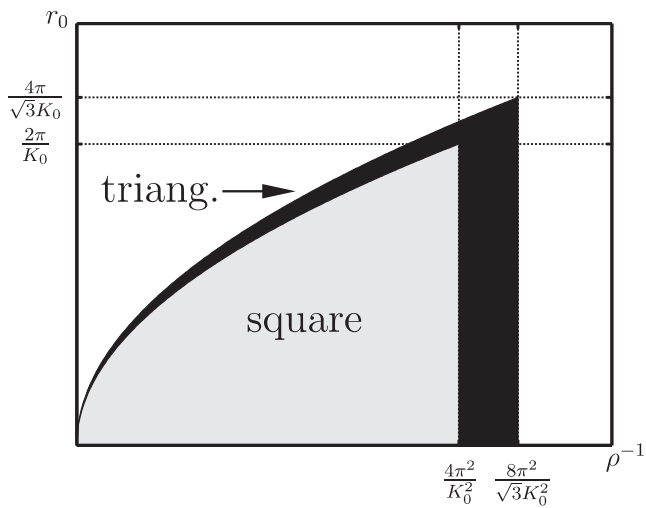}}
\caption{Stability of two-dimensional lattices. The stability intervals of the triangular
and square lattices are obtained as horizontal cuts of the respective domains.}
\end{figure}

The situation in three
dimensions is more complicated, because $\gamma_3=\gamma_{\rm bcc}=\gamma_{\rm fcc}$.
From the expressions
\be\label{bcc-fcc}
r_{\rm bcc}(\rho)=\sqrt{3}/(4\rho)^{1/3}, \qquad r_{\rm fcc}(\rho)=(\sqrt{2}/\rho)^{1/3}
\ee
of the nearest-neighbor distances
one can see that the stability intervals for the bcc and fcc lattices are
\be
I_{\rm bcc}=\left[\frac{K_0^3}{8\sqrt{2}\pi^3},\frac{3\sqrt{3}}{4r_0^3}\right],\quad
I_{\rm fcc}=\left[\frac{K_0^3}{6\sqrt{3}\pi^3},\frac{\sqrt{2}}{r_0^3}\right],
\ee
the lower boundaries being $\rho_3\equiv\rho_{\rm bcc}$ and $\rho_{\rm fcc}$, respectively, cf.
Eq.~(\ref{rhos}). Because at a given density the bcc lattice has the largest $q_{B^*}$
and the fcc the largest $r_B$, the stability interval of any
Bravais lattice is between $\rho_3$ and $\rho_3'=\sqrt{2}/r_0^3$.
According to the argument given above, at $\rho_3$ and $\rho_3'$
the ground state is unique; especially, \emph{at $\rho=\sqrt{2}/r_0^3$ the unique GSC is
the fcc lattice.}
The intervals $I_{\rm bcc}$ and $I_{\rm fcc}$ only partially overlap,
and may not overlap at all: if
$3\pi/2^{1/3}<r_0K_0\leq\sqrt{6}\pi$, $I_{\rm bcc}$ and $I_{\rm fcc}$ are disjoint.
For $r_0<\sqrt{6}\pi/K_0$ the stability intervals of other Bravais lattices fill the gap. At
$r_0=\sqrt{6}\pi/K_0$, $I_{\rm bcc}$ and $I_{\rm fcc}$ shrink to a single point,
$\rho_{\rm bcc}$ and $\rho_{\rm fcc}$, respectively, and for densities in between
no Bravais lattice satisfies
both conditions $q_{B^*}\geq K_0$ and $r_B\geq r_0$. In this interval the ground state is
probably unique and changes continuously from bcc to fcc as the density increases.
In Figure 2 we present the stability intervals of the bcc and the fcc lattices for all
the allowed values of $r_0$.

\begin{figure}[htb]
\centerline{\includegraphics[width=8.5cm]{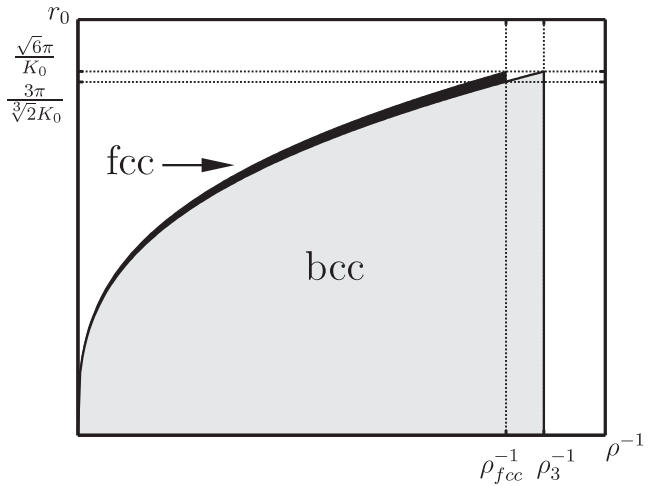}}
\caption{Stability of three-dimensional lattices. The stability intervals of the bcc
and fcc lattices are obtained as horizontal cuts of the respective domains.}
\end{figure}

The GSCs of $\vfi$ will be found by applying
the following extension of the Poisson summation formula.

LEMMA. Let $\fihat\in L^1(\R^d)$ be a function of bounded support, which is
continuous at the origin. Let
$
\vfi(\r)=(2\pi)^{-d}\int\fihat(\k)e^{i\k\cdot\r}\d\k.
$
Choose a Bravais lattice $B$ such that $\fihat$ is continuous at
every $\K$ in $B^*$. Then
\be\label{Psum}
\lim_{\varepsilon\downarrow 0}\sum_{\Rr\in B}e^{-\varepsilon|\r+\Rr|^2}\vfi(\r+\Rr)
=\rho(B)\sum_{\K\in B^*}\fihat(\K)e^{i\K\cdot\r},
\ee
implying the existence of the limit.

Observe that the sum in the right member has only a finite number of nonzero terms, hence
for any $B$ the continuity of $\fihat$ is to be checked only in a finite number of points.
In particular, for any $B$ dense enough the only point of $B^*$ inside the support of $\fihat$
is the origin, where $\fihat$ is continuous. This is precisely the fact we shall use in the
proof of the theorem.

\section{Proofs}

\subsection{Proof of the lemma}

Let
\be
\delta_\varepsilon(\k)=(4\pi\varepsilon)^{-d/2}e^{-k^2/4\varepsilon}.
\ee

First, we show that
\be\label{Peps}
\sum_{\Rr\in B}e^{-\varepsilon|\r+\Rr|^2}\vfi(\r+\Rr)
= \rho(B)\sum_{\K\in B^*}(\fihat*\delta_\varepsilon)(\K)e^{i\K\cdot\r}
\ee
for any $\varepsilon>0$. The argument is essentially the same as the one we used to prove
the Lemma of \cite{Su}.
Because $\vfi(\r)$ is an entire function of $\r$ decaying at infinity,
both $e^{-\varepsilon r^2}\vfi(\r)$ and its Fourier transform $\fihat*\delta_\varepsilon$
are functions of rapid decrease \cite{RS}. Therefore, the infinite sums on both sides of
Eq.~(\ref{Peps}) are absolutely convergent and the convergence is uniform in $\r$. So both
sums define continuous functions, that are periodic with periods $\Rr\in B$ and have the same
Fourier coefficients. Indeed, multiplying Eq.~(\ref{Peps}) by $e^{-i\K\cdot\r}$ and integrating
by terms over the unit cell of volume $\rho(B)^{-1}$,
on the left-hand side after summation we obtain $(\fihat*\delta_\varepsilon)(\K)$,
which is also the trivial result on the right-hand side.
Because of the completeness of the system
$\{e^{i\K\cdot\r}|\K\in B^*\}$ in the Banach space of integrable functions on the unit cell of
$B$, the two continuous periodic functions coincide everywhere.

Next, we prove that the integrability of $\fihat$ and its continuity at $\K$ imply
\be
\lim_{\varepsilon\to 0}(\fihat*\delta_\varepsilon)(\K)=\fihat(\K).
\ee
Fix any $\eta>0$ and write
\bea
\lefteqn{
(\fihat*\delta_\varepsilon)(\K)=\int_{q<\eta}\fihat(\K-\q)\delta_\varepsilon(\q)\d\q
}\nonumber\\
&&+\int_{q>\eta}\fihat(\K-\q)\delta_\varepsilon(\q)\d\q
\equiv
 J_{\varepsilon,<\eta}+J_{\varepsilon,>\eta}.
\eea
If $\varepsilon$ is small enough then $\delta_\varepsilon(\q)<1$ for $q>\eta$, and in
$J_{\varepsilon,>\eta}$ the
integrand can be bounded above by $|\fihat(\K-\q)|$.
Thus, due to the dominated convergence theorem the limit and the integration can be interchanged,
resulting $\lim_{\varepsilon\to 0}J_{\varepsilon,>\eta}=0$, because
$\lim_{\varepsilon\to 0}\delta_\varepsilon(\q)=0$ for $q>\eta$. On the other hand,
\bea
 J_{\varepsilon,<\eta}&=&\fihat(\K)\int_{q<\eta}\delta_\varepsilon(\q)\d\q\nonumber\\
&+&\int_{q<\eta}[\fihat(\K-\q)-\fihat(\K)]\delta_\varepsilon(\q)\d\q.
\eea
Now $\lim_{\varepsilon\to 0}\int_{q<\eta}\delta_\varepsilon(\q)\d\q=1$ and
\bea
\left|\int_{q<\eta}[\fihat(\K-\q)-\fihat(\K)]\delta_\varepsilon(\q)\d\q\right| \nonumber\\
\leq\sup_{q<\eta}|\fihat(\K-\q)-\fihat(\K)|.
\eea
Combining the above equations,
\be
\lim_{\varepsilon\to 0}|(\fihat*\delta_\varepsilon)(\K)-\fihat(\K)|\leq
\sup_{q<\eta}|\fihat(\K-\q)-\fihat(\K)|,
\ee
from which the result follows by letting $\eta$ go to zero.

Finally, using the fact that the different sums involving $\fihat(\K)$ [but not
$(\fihat*\delta_\varepsilon)(\K)$!] are finite, we find
\bea\label{sumdiff}
\lim_{\varepsilon\to 0}\left|\sum_{\K\in B^*}[(\fihat*\delta_\varepsilon)(\K)-\fihat(\K)]
e^{i\K\cdot\r}\right|\nonumber\\
\leq\sum_{\K\in B^*}\sup_{q<\eta}|\fihat(\K-\q)-\fihat(\K)|,
\eea
holding for all $\eta>0$. Thus, we can conclude that the left-hand side is indeed zero.
This completes the proof of the Lemma.

\subsection{Proof of Equation (\ref{max})}

The proof is based on the well-known fact
that $2\pi/q_{B^*}$ is the largest distance between neighboring lattice lines ($d=2$) or
planes ($d=3$) of $B$, see e.g. \cite{AM}. The way we proceed is to consider all the types of
Bravais lattices simultaneously and to select, through a sequence of choices,
the vectors $\a_i$ that define the maximizer of $r_Bq_{B^*}$.
Let $B$ be any two-dimensional Bravais lattice. Choose $\a_1$ among the shortest vectors of $B$,
hence $a_1=r_B$. The lattice line parallel to $\a_1$ is a line of largest density,
therefore the largest distance between neighboring lines is measured perpendicular to $\a_1$.
The other primitive vector, $\a_2$ is selected among the shortest vectors not collinear with
$\a_1$ and making an acute angle $\alpha$ with $\a_1$. Then $\alpha\geq\pi/3$
(otherwise $\a_1-\a_2$ would be shorter, and should replace $\a_2$), and
$2\pi/q_{B^*}=a_2\sin\alpha$ (thus, $b_2=q_{B^*}$).
Therefore
\be
r_Bq_{B^*}=\frac{2\pi a_1}{a_2\sin\alpha}
\ee
whose maximum on the condition that $a_2\geq a_1$ and $\pi/3\leq\alpha\leq\pi/2$ is
$4\pi/\sqrt{3}$,
attained with the choice $a_2=a_1$ and $\alpha=\pi/3$, characteristic to the triangular lattice.

In three dimensions, given $B$, let $P$ be a lattice plane of highest density, containing the
origin. Let $\a_1$ be one of the shortest lattice vectors in the plane, and choose
$\a_2$ among the shortest lattice vectors in $P$ not collinear with $\a_1$ and making an
acute angle $\alpha_{12}$ with it; so we have $a_1\leq a_2$ and, as argued above,
$\pi/3\leq\alpha_{12}\leq\pi/2$.
Because $P$ is of highest density, the largest
distance among lattice planes can be measured perpendicular to it. Accordingly,
$q_{B^*}=b_3$. Choose $\a_3$ among the shortest lattice vectors not contained in $P$
and making an acute angle with at least one of $\a_1$ or $\a_2$.
One of the angles, $\alpha_{13}$ of $\a_1$ and $\a_3$ or $\alpha_{23}$ of $\a_2$ and $\a_3$,
can indeed be obtuse.
However, if it is obtuse, we
replace $B$ by $B^*$ in the line of reasoning and continue with three acute angles:
this will not influence the validity of Eq.~(\ref{max}).
We have to examine two cases. First, suppose that $a_1=r_B$. Then
\bea\label{product}
r_Bq_{B^*}=\frac{2\pi a_1|\a_1\times\a_2|}{|\a_3\cdot(\a_1\times\a_2)|}
=\frac{2\pi a_1}{a_3\sin\alpha}
\eea
where $\alpha$ is the angle of $\a_3$ to $P$.  To maximize $r_Bq_{B^*}$,
we choose $a_3=a_1=r_B$ and then $\alpha$ to be minimum. None of
$\alpha_{13}$ and $\alpha_{23}$ can be smaller than $\alpha_{12}$,
otherwise the density of the plane
spanned by $\a_3$ with either $\a_1$ or $\a_2$ would be higher than that of $P$. Therefore,
the smallest $\alpha$ can be attained if $\a_3$ is in the bisector
plane of $\alpha_{12}$ and
$\alpha_{13}=\alpha_{23}=\alpha_{12}=\pi/3$. But then $a_2$ cannot be larger
than $a_1$, otherwise, again, $P$ was not a plane of
maximum density. Thus, we conclude that $a_1=a_2=a_3=r_B$
and $\alpha_{12}=\alpha_{23}=\alpha_{31}=\pi/3$, specifying the fcc lattice.
The other case is $r_B=a_3<a_1$. Then again $r_Bq_{B^*}=2\pi/\sin\alpha$,
but now $\alpha$ cannot be as small as before, otherwise the density of a lattice plane
containing $\a_3$ would be larger than that of $P$.
Thus, the maximum of $r_Bq_{B^*}$ is indeed attained on the fcc-bcc pair. Its value,
$\sqrt{6}\pi$, is easy to compute.

\subsection{Proof of the theorem, (i)}

Let $X=\cup_{j=1}^J(B_j+\y_j)$, $X_f\subset X$ finite,
and let $R$ be any finite configuration.
If $\fihat$ is continuous at each point of every $B_j^*$ then,
making use of the definitions (\ref{U1})-(\ref{rel}), (\ref{Poisson-Abel}) and
the lemma,
\begin{widetext}
\bea\label{Urel}
U(R|X\setminus X_f)=N_{R}[\fihat(\0)\rho(X)-\vfi(\0)/2]
+\int\fihat(\k)
\left(\left|\sum_{\r\in R}e^{i\k\cdot\r}\,\right|\,^{^2}-
2\sum_{\r\in R}e^{i\k\cdot\r}\sum_{\x\in X_f}e^{-i\k\cdot\x}
\right)\frac{\d\k}{2(2\pi)^{d}}    \nonumber\\
+ \sum_{j=1}^J\rho(B_j)\sum_{\0\neq\K\in B_j^*}\fihat(\K)e^{-i\K\cdot\y_j}
\sum_{\r\in R}e^{i\K\cdot\r}.
\eea
%\end{widetext}
Subtracting the corresponding expression in which $X_f$ replaces $R$, we find
%\begin{widetext}
\bea\label{main}
\lefteqn{
U(R|X\setminus X_f)-\mu N_{R}
-U(X_f|X\setminus X_f)+\mu N_{X_f}
=(N_{X_f}-N_{R})\ [\mu+\vfi(\0)/2-\fihat(\0)\rho(X)]}\nonumber\\
&&
+\int\fihat(\k)
\left|\sum_{\r\in R}e^{i\k\cdot\r}-
\sum_{\x\in X_f}e^{i\k\cdot\x}\,\right|\,^{^2} \ \frac{\d\k}{2(2\pi)^{d}}
+\sum_{j=1}^J\rho(B_j)\sum_{\0\neq\K\in B_j^*}\fihat(\K)e^{-i\K\cdot\y_j}
\left(\sum_{\r\in R}e^{i\K\cdot\r}-\sum_{\x\in X_f}e^{i\K\cdot\x}
\right).\nonumber\\
\eea
\end{widetext}
Suppose now that each $q_{B_j^*}\geq K_0$,
where equality is allowed only if $\fihat(\k)=0$ at $|\k|=K_0$. Any such set of $B_j$\,s satisfies
the continuity assumption. Indeed, $\fihat$ is continuous at $\K=\0$ and, because
$|\K|\geq K_0$ for
every nonzero $\K$ in $\cup_{j=1}^J B_j^*$, $\fihat$ is continuous and takes on
zero in all these
points. Equation (\ref{main}) is, therefore, valid, and the last term on its right-hand side
vanishes for all $R$. The first term can be made zero either by choosing
$N_{R}=N_{X_f}$ or by setting $\mu=\fihat(\0)\rho(X)-\vfi(\0)/2$.
Because $\fihat\geq 0$, these choices prove that $X$ is indeed a GSC and a $\mu$GSC of $\vfi$.

Next, we compute the energy density $e(X)$ of $X$ with each
$q_{B_j^*}\geq K_0$. From Eq.~(\ref{IxX}) and the Lemma,
\be\label{IxX2}
I(\x,X)=\rho(X)\fihat(\0).
\ee
Furthermore, if $Q_L$ denotes the cube of
side $L$ centered at the origin, then
\be\label{limit}
\lim_{L\to\infty}L^{-d}\sum_{\x\in X\cap Q_L}1=\rho(X).
\ee
With Eqs.~(\ref{IxX2}) and (\ref{limit}),
\begin{widetext}
\bea\label{edens}
e(X)&=&\frac{1}{2}\lim_{L\to\infty}L^{-d}\sum_{\x\in X\cap Q_L}\sum_{\x\neq\x'\in X}
\vfi(\x-\x')
=\frac{1}{2}\lim_{L\to\infty}L^{-d}\sum_{\x\in X\cap Q_L}[I(\x,X)-\vfi(\0)]\nonumber\\
&=&-\frac{1}{2}\rho(X)\vfi(\0)+\frac{1}{2}\rho(X)^2\fihat(\0)
=\epsilon\textbf{(}\rho(X)\textbf{)}.
\phantom{a}
\eea
Here the first two equalities define the energy density of a general
configuration, only the third one is specific to a GSC.
We now prove that $\epsilon(\rho)$ is the minimum of $e(X)$
among unions of periodic configurations of density $\rho$. As earlier, we write
an arbitrary (periodic or aperiodic) union of periodic configurations in the form
$
X=\cup_{j=1}^J(B_j+\y_j).
$
Computing $I(\x,X)$ with the help of Eq.~(\ref{Psum}),
inserting it into the definition (\ref{edens})
of $e(X)$ and separating the $\K=\0$ term we find
\bea
e(X)=\epsilon\textbf{(}\rho(X)\textbf{)}
+\frac{1}{2}\sum_{i,j=1}^J\rho(B_i)\sum_{\0\neq \K\in B_i^*}\fihat(\K)
e^{i\K\cdot(\y_j-\y_i)}
\lim_{L\to\infty}L^{-d}\sum_{\Rr\in B_j\cap (Q_L-\y_j)}e^{i\K\cdot\Rr}.
\eea
The limit can be evaluated: it yields $\rho(B_j)$ if $\K\in B_i^*\cap B_j^*$ and zero if
$\K\in B_i^*\setminus B_j^*$.
Hence, we obtain
\bea\label{eunion}
e(X)&=&\epsilon\textbf{(}\rho(X)\textbf{)}
+\frac{1}{2}\sum_{i,j=1}^J\rho(B_i)\rho(B_j)\sum_{\0\neq \K\in B_i^*\cap B_j^*}\fihat(\K)
e^{i\K\cdot(\y_j-\y_i)}
\nonumber\\
&=&\epsilon\textbf{(}\rho(X)\textbf{)}+\frac{1}{2}\sum_{\0\neq \K\in \cup B_i^*}\fihat(\K)
\left|\sum_{j=1}^J\chi_{B_j^*}(\K)\rho(B_j)e^{i\K\cdot\y_j}\right|^2
\eea
\end{widetext}
where $\chi_{B_j^*}(\K)=1$ if $\K$ is in $B_j^*$ and is zero otherwise. Thus,
$e(X)\geq\epsilon\textbf{(}\rho(X)\textbf{)}$, as claimed.

\subsection{Proof of the theorem, (ii): absence of metastability}

The following proposition makes no use of Fourier transforms, but exploits directly the
summability assumptions (\ref{abssum}) and (\ref{limsum}).

PROPOSITION. (i) Let $\vfi$ be a strongly tempered pair potential. Let $X$ and $Y$
be locally uniformly
finite configurations such that the limits (\ref{limit}) and (\ref{edens}), defining the density
and the energy density, exist. If $\rho(X)=\rho(Y)$ but
$
e(X)>e(Y),
$
then $X$ is not a GSC of $\vfi$, that is,
$X$ does not satisfy the local stability condition (\ref{GSC'}) restricted to
number-preserving perturbations. (ii) If $\vfi$ is not strongly tempered but summable on Bravais
lattices in the sense of Eq.~(\ref{limsum}), and $X$ and $Y$ are unions of periodic configurations
with $\rho(X)=\rho(Y)$ and
$
e(X)>e(Y),
$
then $X$ is not a GSC of $\vfi$.

\emph{Proof.}| The proof was already outlined in \cite{Su}. Here we give the missing details.
(i) In the case of strongly tempered interactions
consider the cubes $Q_L$ introduced above. In general,
$\Delta_L=N_{X\cap Q_{L}}-N_{Y\cap Q_{L}}\neq 0$ but, because $\rho(X)=\rho(Y)$,
$\Delta_L=o(L^d)$.
Let $Y_L$ be a configuration in $Q_L$ obtained from $Y\cap Q_L$ by adding or deleting
$|\Delta_L|$ points so that $N_{Y_L}=N_{X\cap Q_L}$.
From the definitions (\ref{rel}) and (\ref{edens}) we deduce
\be\label{U1'}
U(X\cap Q_L|X\setminus Q_L)=e(X)L^d+\frac{1}{2}I(X\cap Q_L,X\setminus Q_L)+o(L^d).
\ee
On the other hand, by strong temperedness
\[
U(Y_L|X\setminus Q_L)=U(Y\cap Q_L|X\setminus Q_L)+o(L^{d}),
\]
and therefore
\begin{widetext}
\bea\label{U2}
U(Y_L|X\setminus Q_L)=e(Y)L^d+I(Y\cap Q_L,X\setminus Q_L)
-\frac{1}{2}I(Y\cap Q_L,Y\setminus Q_L)+o(L^d).
\eea
To conclude that
\bea
U(Y_L|X\setminus Q_L)-U(X\cap Q_L|X\setminus Q_L)\asymp [e(Y)-e(X)]L^d
<0\nonumber
\eea
for $L$ large enough, we have to show that the $I$ terms in Eqs.~(\ref{U1'}) and (\ref{U2})
are of smaller order, $o(L^{d})$. Again, this holds because $\vfi$ is strongly tempered.
For example,
\bea
L^{-d}|I(X\cap Q_L,X\setminus Q_L)|&=&
L^{-d}|I(X\cap Q_{L-\sqrt{L}},X\setminus Q_L)+
I(X\cap Q_L\setminus Q_{L-\sqrt{L}}\,,X\setminus Q_L)|\nonumber\\
&\leq& \rho(X)\left[\sup_\r\sum_{\x\in X, |\x-\r|\geq\sqrt{L}}|\vfi(\x-\r)|
+L^{-d/2}\sup_\r\sum_{\x\in X}|\vfi(\x-\r)|\right]+o(1)
\eea
\end{widetext}
which indeed tends to zero as $L$ goes to infinity.
(ii) In the case when $\vfi$ is not strongly tempered, let $X=\cup_{j=1}^J(B_j+\y_j)$ and
$Y=\cup_{j=1}^{J'}(B'_j+\y'_j)$, with primitive vectors $\a_{j,\alpha}$ and $\a'_{j,\alpha}$,
respectively. Consider the set
$\cup_{j=1}^J\{\sum_\alpha n_\alpha\a_{j,\alpha}+\y_j|\ |n_\alpha|<N_\alpha\}$,
and let $X_f$ be the
union of $J'$ non-overlapping adjacent translates of this set. Similarly, let $R$ be the union
of $J$ non-overlapping adjacent translates of the set
$\cup_{j=1}^{J'}\{\sum_\alpha n_\alpha\a'_{j,\alpha}+\y'_j|\ |n_\alpha|<N_\alpha\}$.
In this way $N_R=N_{X_f}$. Defining $V_R$ and $V_{X_f}$ by the equalities $U(R)=e(Y)V_R$ and
$U(X_f)=e(X)V_{X_f}$, $V_R/V_{X_f}\to 1$ as all $N_\alpha\to\infty$, because $\rho(X)=\rho(Y)$.
On the other hand, from the convergence (\ref{limsum}) it follows that the interaction terms are
of smaller order, $V_{R}^{-1}I(X_f,X\setminus X_f)$ and $V_{R}^{-1}I(R,X\setminus X_f)$ tend to
zero. This ends the proof of the proposition.

The proposition implies the assertion (ii) of the Theorem,
because any ground state configuration
$Y$ of density $\rho(Y)=\rho(X)$, described in (i), has an energy
density $e(Y)=\epsilon\textbf{(}\rho(X)\textbf{)}<e(X)$.

\section{Summary}

In this paper we have described the ground state configurations of pair interactions $\vfi$ and
$\vfi+\psi$, where $\vfi$ is the inverse Fourier transform of a nonnegative function vanishing
outside the sphere of radius $K_0$, and $\psi$ is a nonnegative finite-range interaction of range
$r_0$.
The GSCs of $\vfi$ alone could be obtained, as already in \cite{Su},
above a threshold density $\rho_d$, while those of $\vfi+\psi$ in a density interval whose
lower boundary is $\rho_d$ and the upper boundary is the close-packing density of hard balls of
diameter $r_0$. Below this density, $\psi$ alone would allow as a GSC any configuration
with a nearest-neighbor distance not smaller than $r_0$; combined with
$\vfi$, its role is to decrease the degeneracy of the GSCs of $\vfi$. This reduced degeneracy
is still continuous inside the interval, but at the boundaries $\vfi+\psi$ has a unique GSC,
which is, in three dimensions, the bcc lattice at the lower and the fcc lattice at the upper
density limit.
This transition from a bcc ground state at $\rho=\rho_3$ to an fcc one at
$\rho=\sqrt{2}/r_0^3$ is the most
interesting finding of the present work.

The method used in \cite{Su} and in this paper cannot be applied to obtain the ground states
of $\vfi$ below $\rho_d$.
The threshold value is a true critical density separating the high-density
continuously degenerate region from the low-density region in which the GSC is
presumably unique, apart from Euclidean transformations.
If $\fihat(\0)>0$, the analytic form of the relation between the density and the chemical
potential also changes at $\rho_d$ \cite{Su}.
We expect that at least in a subclass of interactions
the unique bcc ground state at $\rho_3$ survives at lower densities.
Similarly, the mathematical method used here is not suitable to obtain the ground
states of $\vfi+\psi$ above the upper density limit. Again, this limit seems to be a true
critical value,
with a continuously degenerate ground state below and, probably, a unique ground state above it.
This unique ground state may depend on the details of $\psi$, but we expect it to be the fcc
lattice for a subclass of positive interactions. To clarify these questions, and also the
nature of the curious liquid-like ground state between the fcc and bcc phases will be the
subject of future research.

This work was supported by OTKA Grants T 043494 and T 046129.

\end{document}